# A Novel General Theoretical Derivation of Newton's Experimental Formula for Collision of Particles

R.C.Gupta*, Ruchi Gupta** and Sanjay Gupta***

An experimental formula, sometimes named as Newton's collision-formula, $(v_1-v_2) = -e.(u_1-u_2)$ relating relative-velocities before & after impact of two bodies under linear-collision, is commonly used successfully for study of collision-dynamics. Although it seems possible as shown in text-books [1,2] to derive this relation, assuming (defining) 'e' as ratio of impulses during restitution & deformation; but as yet, neither a good theoretical-basis nor a general-derivation of this formula based on obvious choice of energy-consideration, has been reported. In this brief-paper a relativistic-theoretical-basis is adopted and a general formula-derivation is given. It is shown, interestingly, that proof of the Newton's collision-formula is hidden in Einstein's-special-relativity.

Consider two particles 1 & 2 moving in the same (+ive x) direction with initial velocities $u_1$ & $u_2$ before impact and that its velocities after the impact are $v_1$ & $v_2$ in the same directions; such that relative-velocity before-impact is $u_r$ and that after-impact it is $v_r$. Newton's famous experimental collision-formula given in text-books [1,2] relates these with the equation $v_r/u_r = -e$ or $(v_1-v_2)/(u_1-u_2) = -e$, where e is said to be coefficient of restitution ( normally, $0 < e < 1$; but for perfect-elastic-collision $e = 1$ and for perfect-plastic-collision $e = 0$). The relativistic basis and a novel general theoretical derivation of this formula based on energy-consideration are described as follows.

In an effort to derive this formula a relativistic approach / basis is chosen. It may be noted that according to Einstein's 'special theory of relativity' [3] conclusion made by a moving-observer in his-frame of reference is same as the conclusion made by a laboratory-observer in the laboratory-frame. Now, imagine a mass-less observer sitting on particle-2 (or the particle-2 itself as an observer) so that to him - particle-2 is at rest and particle-1 moves with a relative-velocity $u_r$ before-impact and with relative-velocity $v_r$ after-impact. Mass-energy ($E = mc^2$) conservation before & after impact yields the following equation, using the relativistic mass $m = m_o/(1-v^2/c^2)^{1/2}$; where $m_{1o}$ & $m_{2o}$ are the rest-masses of the two particles with-respect-to the moving observer, c is velocity of light.

$$m_{1o}/(1-u_r^2/c^2)^{1/2} \cdot c^2 + m_{2o}.c^2 = m_{1o}/(1-v_r^2/c^2)^{1/2}.c^2 + m_{2o}.c^2 + \text{losses} \qquad (1)$$

This equation reduces to the following equation, neglecting higher-order-terms (in binomial-expansion) since collision-velocities are much smaller than the velocity of light.

$$\tfrac{1}{2} m_{1o}.u_r^2 = \tfrac{1}{2} m_{1o}.v_r^2 + \text{losses} \qquad (2)$$

Considering that energy-loss in collision is a fraction ($\delta$) of the energy $\tfrac{1}{2} m_{1o}.u_r^2$, this equation further reduces to,

$$v_r / u_r = \pm (1-\delta)^{1/2} \qquad (3)$$

---

* Dr. R.C.Gupta, Professor & Head, Dept. of Mechanical Engineering, Institute of Engineering & Technology (IET), Lucknow, India. ( E-mail: rcg_iet(at)homail.com ).
** Ruchi Gupta, Software Engineer, Cisco Systems, CA, USA. ( ruchig(at)stanfordalumni.org ) .
*** Sanjay Gupta, Staff Engineer, Vitria Technologies, CA, USA. ( sanjaygu(at)gmail.com ) .

Since $u_r$ is the relative approaching velocity before-impact and that for-collision the two particles approach to the closest, and since after-collision these have to separate out; +ive sign on RHS of the equation-3 is omitted and –ive sign is retained. Taking (defining) $e = (1-\delta)^{1/2}$, the Newton's experimental collision-formula is arrived at as following (Eqs. 4 & 5). It may be noted (from $e = (1-\delta)^{1/2}$) that for perfect-elastic-collision with no energy loss $\delta = 0$ hence $e = 1$, and that for perfect-plastic-collision with full energy loss $\delta = 1$ hence $e = 0$.

$$v_r / u_r = -e = (1-\delta)^{1/2} \qquad (4)$$

Further, from Einstein's relativistic velocity addition [3] it is known that $u_r = (u_1-u_2)/(1+u_1.u_2/c^2)$ and $v_r = (v_1-v_2)/(1+v_1.v_2/c^2)$ which are approximately simplified (as per Galilean relativity) as $u_r = (u_1-u_2)$ and $v_r = (v_1-v_2)$; hence the derived relation (Eq.4) $v_r / u_r = -e$ gives the following commonly known/used Newton's experimental collision- formula (Eq.5).

$$(v_1 - v_2) / (u_1-u_2) = -e \qquad (5)$$

Thus the novel theoretical relativistic derivation of Newton's experimental formula for collision of two particles has been made, although using some approximations (between Eq.1 & Eq.2 and between Eq.4 & 5)); which means that Newton's formula is not exact and can not be used for very high-velocity-collisions (of elementary particles). In that situation, the following relativistic-velocity-added modified formula (Eq.6) could possibly be used which too, however, is approximate but more accurate than the Newton's formula (Eq.5).

$$[ (v_1-v_2) / (u_1-u_2) ] . [ (1 + u_1.u_2/c^2) / (1 + v_1.v_2/c^2) ] = -e \qquad (6)$$

It is interesting to note that the theoretical-basis of the *Newton's* collision-formula lies in the *Einstein's* special-relativity.

In an interesting paper [4] it is shown that the second-law-of-thermodynamics and the special-relativity are linked together. Thus it seems that differently-looking entities such as the second-law-of-thermodynamics (engine-efficiency η, $0 < η < 1$), the special-relativity (velocity-ratio β, $0 < β < 1$), and the collision-dynamics (restitution-coefficient e, $0 < e < 1$) are in fact not-so-different but in a way the different facets of the same coin. Also, it is well known [3] that special-relativity is the 'bridge' between electricity and magnetism. In a recent fundamental paper [5] it is shown that once-again it is the special-relativity which is the 'key' for much-sought *unification* of gravity to electromagnetic-force.